\documentclass[preprint,aps]{revtex4}
\usepackage{hyperref,color}
\usepackage{amssymb}
\usepackage{graphicx}
\usepackage{epstopdf}

\begin{document}
\title{Survival of New Physics: An Anomaly-free Neutral Gauge Boson at the LHC}
\author{Ying Zhang$^1$\footnote{{\it Email address}: hepzhy@mail.xjtu.edu.cn},
Qing Wang$^{2,3}$\footnote{{\it Email address}: wangq@mail.tsinghua.edu.cn.}}

\address{$^1$School of Science, Xi'an Jiaotong University, Xi'an, 710049, P.R.China\\
$^2$Department of Physics, Tsinghua University, Beijing 100084, P.R.China \\
$^3$Center for High Energy Physics, Tsinghua University, Beijing 100084, P.R.China}
\date{\today}

\begin{abstract}
An anomaly-free $U(1)'$ effective Lagrangian as the most common new physics beyond the standard model is proposed to survey the maximal parameter space constrained by electroweak precise measurements at the LEP and direct detection in dilepton decay channel at the LHC at $\sqrt{s}=7$ TeV. By the global fit of effective couplings of $Z$ boson to the SM fermions, $\Delta_{11},\Delta_{21},g_2\Delta_{31}$ related to mixings and $r$ related to $U(1)'$ charge assignment are bounded. The allowed areas are plotted in the not only $r$-$g_2$ but also $m_{Z'}$-$g_2$ planes, which show that a sub-TeV $Z'$ is still permissible as long as the coupling $g_2\sim 0.01$. The results provides a prime requirement to an extra $U(1)'$ gauge boson and hinds the direction of possible new physics beyond the standard model. The possible signal in dilepton decay channel at LHC at $\sqrt{s}=14$ TeV is also provided.
\end{abstract}
\maketitle
\section{motivation}

When people were exciting for the found of Higgs-like particle with about 125GeV mass at the LHC,
we have to worry about the survival of new physics (NP) right away. Almost all experiments are proving the Standard Model (SM), the space left to NP is less and less. There is an impending question we are dying to know that how much space can NP survive.
It is a good checking point to choose a neutral gauge boson, which often appears in GUT and superstring model associated with $U(1)'$ group, as a popular candidate of NP beyond the SM.
	
	There are many relative issues summarized by A. Leike and P. Langacker \cite{LeikeReview,LangackerReview}. However, duo to different motivations, $Z'$ interactions to the SM fermions is set by models, which makes the results are highly model-dependent. The minimal mass of the vector boson is limited at about $1.8$ TeV. In the paper, we relax any possible motivations and roles coming from underlying theory or phenomenology, and construct a model-independent effective Lagrangian to describe $U(1)'$ gauge boson $Z'$ only following the requirement of gauge symmetry. In bosonic sector, all possible mass and kinetic mixings meeting gauge symmetry are investigated. And in fermionic sector, anomaly-free charge assignments is required to satisfy gauge symmetry. The interactions to fermions are dominated by a global coupling $g_2$ and a charge assignment parameter $r$. They are keys to realize model-independent description. We consider constrains from not only electroweak precise observables but also direct detection at LHC, and then decide the possible parameter space left to the simplest NP particle.
	
	The article is organized as following: firstly,  anomaly-free $Z'$ effective Lagrangian is constructed. And then, we diagonalize gauge bosons mixing matrixes to obtain mass eigenvalues of neutral bosons. The limit to parameters is studied based on electroweak precise measurements in LEP and direct search at the LHC at $\sqrt{s}=7$ TeV. The allowed area is shown in $m_{Z'}$-$g_2$ plane. Finally, the possible signal in dilepton final states at $\sqrt{s}=14$ TeV at the LHC is predicted.	
\section{effective Lagrangian}

To construct-independent $Z'$ effective theory, we will focus on parameterizations in $Z'$ mixings and interactions. Denote $U(1)'$ gauge eigenstate as $X_\mu$. On gauge eigenstates base $(W_\mu^3,B_\mu,X_\mu)^T$, the covariant derivative has the form
  $$D_\mu \hat{U}=\partial_\mu\hat{U}+igW_\mu\hat{U}-ig'\hat{U}\frac{\tau_3}{2}B_\mu-i{g''}\hat{U}X_\mu.$$
Here, $\hat{U}$ is non-linear realization of Goldstone bosons. $W_\mu$ and $B_\mu$ are respectively gauge field of $SU(2)_L$ and $U(1)_Y$ with gauge coupling $g$ and $g'$.
Using $SU(2)_L$ covariant building blocks $T\equiv \hat{U} \tau_3 \hat{U}^\dag$ and $V_\mu\equiv (D_\mu \hat{U})\hat{U}^\dag$, mass terms arise from 4 operators in $p^2$ order: $tr[V_\mu V^\mu],tr[TV_\mu]^2,tr[V_\mu]^2$ and $tr[TV_\mu]tr[V^\mu]$.
The first operator corresponds to the electroweak standard model.
The second one provides an extra mass correction to isospin third-component $W^3_\mu$.
The third one generates non-standard mixing between $B_\mu$ and $W^3_\mu$.
And the last one parameterizes $Z$-$Z'$ mixing. However, the second and third one can be absorbed in the re-definition of gauge couplings $g$ and $g'$, and are not independent \cite{OurCPC2012}.
Similarly, kinetic mixing terms are also controlled by 4 operators: $tr[TW_{\mu\nu}]^2$, $tr[TW_{\mu\nu}]B^{\mu\nu}$, $tr[TW_{\mu\nu}]X^{\mu\nu}$ and $B_{\mu\nu}X^{\mu\nu}$ \cite{OurJHEP2008,OurJHEP2009}.
The first operator corresponds a correction to $W_\mu^3$ kinetic term. The second one yields kinetic mixing between $W^3_\mu$ and $B_\mu$. The third and forth ones cause $U(1)'$ boson $X_\mu$ kinetic mixings with $W^3_\mu$ and $B_\mu$, respectively. The first two operators are non-standard term beyond the SM and there is no any reason to neglect these invariant ones \cite{EWCL}.

Expressing these operators in obvious gauge fields, Lagrangian related to mixings is written as
\begin{eqnarray*}
\mathcal{L}_{mix}
&=&\frac{m_0^2}{2}(c_W W_\mu^3-s_W B_\mu)^2
	+\frac{m_{1}^2}{2}X_\mu X^\mu
	+2\beta m_{0} m_{1}X_\mu(c_W W_\mu^3-s_W B_\mu)
	\\
		&&-\frac{1}{4}B_{\mu\nu}B^{\mu\nu}-\frac{1}{4}X_{\mu\nu}X^{\mu\nu}
		-\frac{1}{4}(1-\alpha_b)(W_{\mu\nu}^3)^2
		\\
		&&+\frac{1}{2}\alpha_a B_{\mu\nu}W_{\mu\nu}^3
		+\alpha_{c}X^{\mu\nu}W_{\mu\nu}^3
		+\alpha_{d}X^{\mu\nu}B_{\mu\nu}
	\end{eqnarray*}
with mass mixing $\beta$ \cite{OurCPC2012}. Here, $m_0$ and $m_1$ are $Z$ and $Z'$ mass in gauge eigenstates and $s_W,c_W$ are sine and cosine of Weinberg angle.

To $Z'$ interactions to EW gauge bosons $W^\mu$ and $Z$, some independent parameters control them, which may come from underlying theory and not vanish even though no $Z'$ mixings.
For example the decay channel $\Gamma(Z'\rightarrow W^+W^-)$ may arise from $Z$-$Z'$ mixing or high order operator $X_{\mu\nu}tr[T[V^\mu,V^\nu]]$. The former is suppressed by small $Z$-$Z'$ mixing angle, and the later stands for possible NP which has no any promption from high energy experiments. For the similar reason, all these decays of $Z'$ to EW bosons are expected to be small.

The extra neutral current interaction is introduced like
	$$\mathcal{L}_{int}=-g_2\sum_f\bar{f}\gamma_\mu(y'_{Lf} P_L+{y'_{Rf}}P_R)f X^{\mu}$$
with left-handed (right-handed) fermionic $U(1)'$ charge $y'_{Lf}$ ($y'_{Rf}$).
To keep gauge symmetry, $U(1)'$ charge assignments must be anomaly-free. For the family universal case, there are 6 independent charges: $y'_l$ and $y'_q$ for left-handed leptons and quarks, $y'_u,y'_d,y'_\nu,y'_e$ for right-handed up-quark, down-quark, neutrino and electron, respectively.
From $[SU(3)_C]^2U(1)'$,
$[SU(2)]_L^2U(1)'$ and $[U(1)_Y]^2U(1)'$ cancellation requirements, we have
 \begin{eqnarray*}
	y'_l=-3y'_q,~~~
	y'_d=2y'_q-y'_u,~~~
	y'_e=-2y'_q-y'_u.
 \end{eqnarray*}
$U(1)_Y[U(1)']^2$ anomaly is cancelled automatically.
If and only if the number of right-handed neutrinos is 3, $[U(1)']^3$ anomaly and  gravitational-gauge mixing anomaly can be satisfied simultaneously and the charge is $y'_{\nu}=-4y'_q+y'_u$. Without loss of generality, the coupling $g_2$ is normalized so that $y'_{u}=1$. Now,
these couplings are dominated by two free parameters: coupling $g_2$ that controls global intensity and charge ratio $r(\equiv y'_q/y'_u)$ that assigns flavor charges
\begin{eqnarray*}
y'_l=-3r,~,y'_q=1,~y'_u=1,~y'_d=2r-1,~y'_e=-2r-1,~y'_\nu=-4r+1.
\end{eqnarray*}

Briefly, anomaly-free $Z'$ effective theory, inspired by $U(1)'$ gauge symmetry, can be parameterized by mass mixing $\beta$, kinetic mixing $\alpha_{c},\alpha_d$ in bosonic sector ($\alpha_a$ and $\alpha_b$ respectively parameterize non-standard $W^3_{\mu\nu}W^{3\mu\nu}$ and $W^3_{\mu\nu}B^{\mu\nu}$ electroweak bososic kinetic mixing terms, which are not relative to $Z'$ boson), coupling $g_2$ and the charge ratio $r$ in fermionic sector. The full Lagrangian for $U(1)'$ boson is
	$$\mathcal{L}_{Z'}=\mathcal{L}_{mix}+\mathcal{L}_{int}.$$
\section{diagonalization matrix}
To obtain mass eigenstates, let's make a rotation $U$ 	\begin{eqnarray}
	\left(\begin{array}{c} W^3_\mu\\ B_\mu\\ X_\mu\end{array}\right)=U\left(\begin{array}{c} Z_\mu\\ A_\mu\\ Z'_\mu\end{array}\right)
	\label{Urotation}
	\end{eqnarray}
to diagonalize mass and kinetic mixings simultaneously.
Considering rotation $U$ reducing to Weinberg's rotation when no $Z'$ mixings, $U$ may be expressed as the sum of Weinberg's rotation and $Z'$ mixing corrections
\begin{eqnarray*}
	U=\left(\begin{array}{ccc}
		c_W+\Delta_{11}& s_W+\Delta_{12} &  \Delta_{13}
		\\
		-s_W+\Delta_{21} & c_W+\Delta_{12} & \Delta_{23}
		\\
		\Delta_{31} & \Delta_{32} &1+ \Delta_{33}
		\end{array}\right).\label{Umatrix}
\end{eqnarray*}
Notice that 9 $\Delta_{ij}$ ($i=1..3,j=1..3$) are not independent, they are determined by only 7 phenomenological parameters: mass $m_1$ and $m_2$, mass mixing $\beta$, and kinetic mixings $\alpha_{a,b,c,d}$, i.e. $\Delta_{ij}=\Delta_{ij}(m_1,m_2,\beta,\alpha_a,\alpha_b,\alpha_c,\alpha_d)$. So, we must find two constraint conditions on $\Delta$s. One is $s_W\Delta_{22}=c_W\Delta_{12}$,
which results from the requirement for a massless photon.
Another constraint is $\Delta_{32}=0$ due to the requirement of keeping photon coupling vector-type \cite{StueckelbergNote}.

After the rotation (\ref{Umatrix}), the mass eigenvalues of $Z$ and $Z'$ are
	\begin{eqnarray*}
		m_Z^2&=&m_0^2(1+2c_W\Delta_{11}-2s_W\Delta_{21})+4\beta m_0m_1\Delta_{31}+\mathcal{O}(\Delta^2)
		\\
		m_{Z'}^2
		&=&m_1^2(1+2\Delta_{33})+4\beta m_0m_1(c_W\Delta_{13}-s_W\Delta_{23})+\mathcal{O}(\Delta^2).
	\end{eqnarray*}

\section{constraints in LEP}
Due to the good fit of the SM to electroweak precise observables, $Z'$ effective theory must be constrained by precise electroweak experiments. The heavy neutral boson contributes to low energy observables by two fashions: mixings and $Z'$ exchange. They often coexist in phenomenology.
However, corrections to fermionic couplings only come from mixings.
A theoretical observables can be divided into the SM part and $Z'$ contribution
	$$\mathcal{O}_{th}=\mathcal{O}_{SM}+\Delta\mathcal{O}_{Z'}.$$
Due to the triumph of the SM, $\Delta\mathcal{O}_{Z'}$ must be very small. Generally, it's a safe precession to neglect high order effect of $Z'$ \cite{LeikeReview}. Using rotation (\ref{Urotation}), vector and axial-vector couplings in weak neutral current are given by
	\begin{eqnarray*}
	g^f_{V,A}=g^{f,0}_{V,A}+\Delta g^f_{V,A}
	\end{eqnarray*}
$g^{f,0}_{V,A}$ is the SM couplings in tree level, $g^{f,0}_{V}=t_{3L}^f-2q^fs_W^2$ and $g^{f,0}_{A}=t_{3L}^f$. The effective couplings with radiative corrections to propagators and vertices in the SM can be found in LEP report \cite{LEP2006}. The $Z'$ corrections are
	\begin{eqnarray*}
		\Delta g^f_{V,A}=c_W\Delta_{11} t_{3iL}+s_W\Delta_{21}(y_{iL}\pm y_{iR})+\frac{s_Wc_W}{e}g_2\Delta_{31}(y'_{iL}\pm y'_{iR}).
	\end{eqnarray*}
$Z'$ corrections are constrained by $Z$-pole observables. The validity of constraint depends on two factors: experiments precision and calculation precision based on the SM. Although there are 14 observables in LEP/SLD at $Z$-pole, the SM calculation can be parameterized into only 4 radiate correction factors: $\Delta\rho,\Delta r_W, \Delta\kappa$ and $\Delta\rho_b$ (or express into 4 new parameters $\epsilon_{1},\epsilon_{2},\epsilon_3,\epsilon_b$ introduced by Altarelli, et.al. \cite{LEP2006,EWepsilon}).
Considering the number of independent measurements in experiment and the SM calculation, it's a {\it balanced treatment} to choice pseudo observables, 8 effective coupling constants $g_{V,A}^f$ for $f=l,\nu,c,b$, to limit $Z'$ parameters. We minimize
	\begin{eqnarray*}
		\chi^2=\sum_{f}\frac{(g_{V,A}^{f,exp}-g_{V,A}^{f,SM}-\Delta g_{V,A}^{f})^2}{(\delta g_{V,A}^{f,exp})^2}
	\end{eqnarray*}
where supercript $^{exp}$, $^{SM}$ respectively denote the corresponding experiment values and the SM fit values,  and $\delta g_{V,A}^{f,exp}$ are their experimental errors.
The four free parameters are $\Delta_{11},\Delta_{21},g_2\Delta_{31}$ and $r$.
As we have mentioned, $\Delta_{ij}$ are the functions of mixing parameters. We must keep the fit parameters independent. It can be proved by calculating rotation matrixes invoked by mass mixing $\beta$, and kinetic mixings $\alpha_{c,d}$, respectively, even if EW boson kinetic mixing $\alpha_{a}$ and $\alpha_b$ vanishing.
After detailed calculations, we arrive at the global fit results in Table. \ref{Tab.fitresults}.
$Z'$ slight improves fit confidence level from 93\% (about $1.8\sigma$) to 96\% (about $2.1\sigma$).
The parameter ranges are shown in Table. \ref{Tab.range}.
\begin{table}[htdp]
\caption{LEP experiment results on the effective coupling constants and the SM Z-pole fit. Data come from Table 7.9  and Table G.3 in literature \cite{LEP2006}. The last  row represents the corresponding C.L..}
\begin{center}
\begin{tabular}{c|c|c|c}
\hline\hline
coupling & exp. & SM fit & $Z'$ fit
\\
\hline
$g_A^\nu$ & $+0.50075\pm0.00077$ & $+0.50199\pm^{0.00017}_{0.00020}$  & $+0.50063$
\\
$g_A^l$ & $-0.50125\pm0.00026$ & $-0.50127\pm^{0.00020}_{0.00017}$   & $-0.50116$
\\
$g_A^b$ & $-0.5144\pm0.0051$ & $-0.49856\pm^{0.00041}_{0.00020}$   & $-0.49845$
\\
$g_A^c$ & $+0.5034\pm0.0053$ & $+0.50144\pm^{0.00017}_{0.00020}$   & $+0.50013$
\\
\hline
$g_V^\nu$ & $+0.50075\pm0.00077$ & $+0.50199\pm^{0.00017}_{0.00020}$    & $+0.50063$
\\
$g_V^l$ & $-0.03753\pm0.00037$ & $-0.03712\pm0.00032$   & $-0.03751$
\\
$g_V^b$ & $-0.3220\pm0.0077$ & $-0.34372\pm^{0.00049}_{0.00028}$   & $-0.34267$
\\
$g_V^c$ & $+0.1873\pm0.0070$ & $+0.19204\pm0.00023$   & $+0.19185$
\\
\hline
$\chi^2$/dof & - & $24.6/8$ & $20.1/8$
\\
P value & - & $93\%$ & $96\%$
\\
\hline\hline
\end{tabular}
\end{center}
\label{Tab.fitresults}
\end{table}
\begin{table}[htdp]
\caption{global fit results. The corresponding errors come from diagonal elements of the inverse of Hessian matrix. The ranges in $2\sigma$ C.L. are listed in the last column.}
\begin{center}
\begin{tabular}{c|c|c|c}
\hline\hline
quantity & fit result & range in $2\sigma$
\\
\hline
$\Delta_{11}$ & $-0.00067\pm0.00040$ & $(-0.00147,0.00013)$
\\
$\Delta_{21}$ & $0.0017\pm0.0076$ & $(-0.0135,0.0169)$
\\
$g_2\Delta_{31}$ & $-0.00044\pm0.0018$ & $(-0.00404,0.00316)$
\\
$r$ & $-0.015\pm1.1$ & $(-2.215,2.185)$
\\
\hline\hline
\end{tabular}
\end{center}
\label{Tab.range}
\end{table}%
Notice that $Z'$ mass does not been limited by effective couplings. Generally, $m_{Z'}$ can be limited by $\rho$ or $Z$ mass correction by $Z-Z'$ mixing. There are enough more results on the issue in literature \cite{zprimebound} which shown that a small mass mixing corresponds a heavy $Z'$ and vice versa.
For the typical value $\beta\sim 10^{-3}$, $m_{Z'}$ is several TeV.
\section{search at the LHC}	

\begin{figure}[htbp]
	\begin{center}
	\includegraphics[scale=0.10]{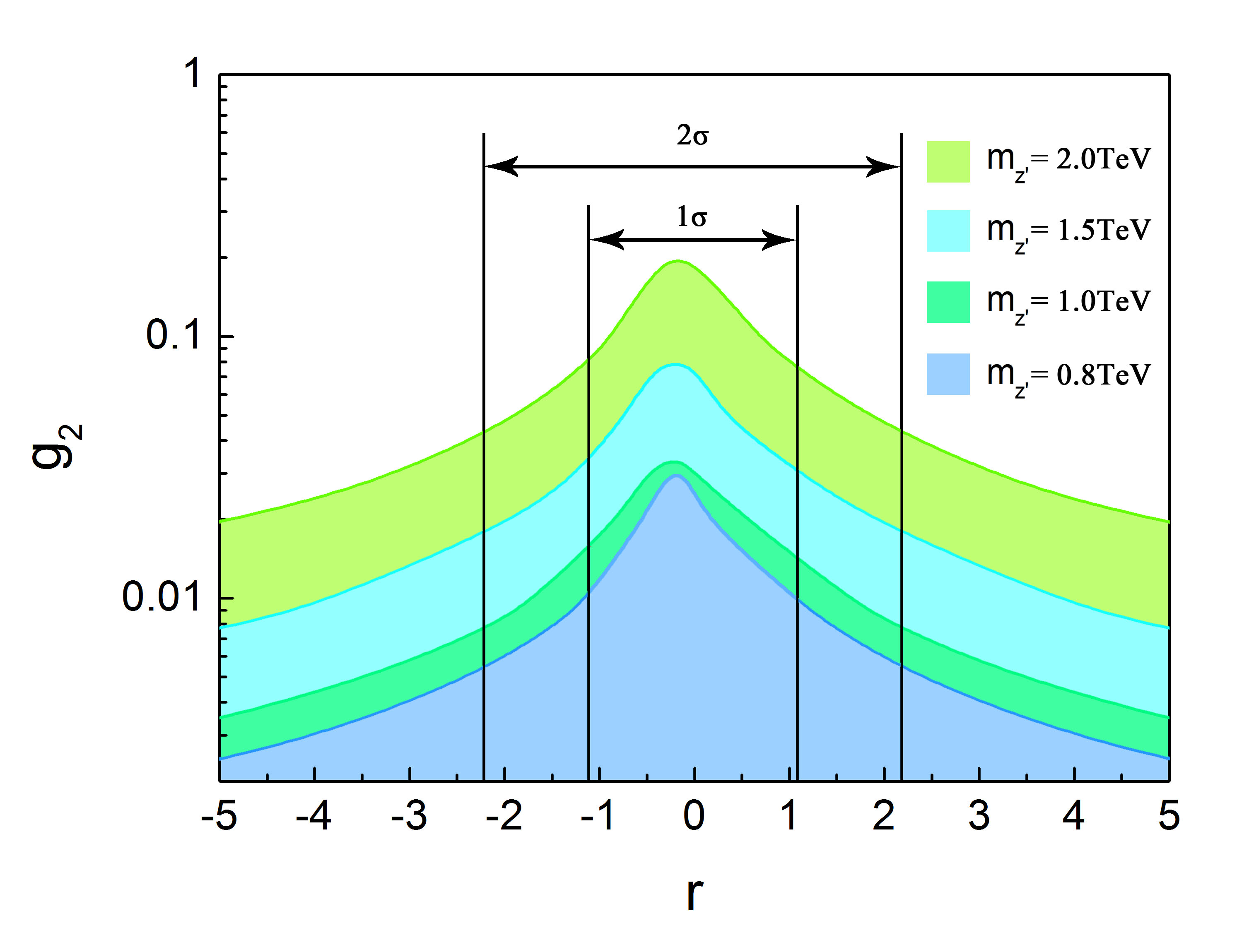}
	\caption{$95\%$ C.L. possible allowed area in the $g_2$ vs $r$ plane at 7 TeV LHC. Exclusion lines correspond to excepted $2\sigma$  signal regions at ATLAS in dilepton decay channel in Fig.3 of \cite{ATLASZprime}. Black vertical lines denote $r$ fit ranges in $1\sigma$ and $2\sigma$ C.L., respectively.}
	\label{Fig.ATLASlimit}
	\end{center}
	\end{figure}

\begin{figure}[htbp]
	\begin{center}
	\includegraphics[scale=0.09]{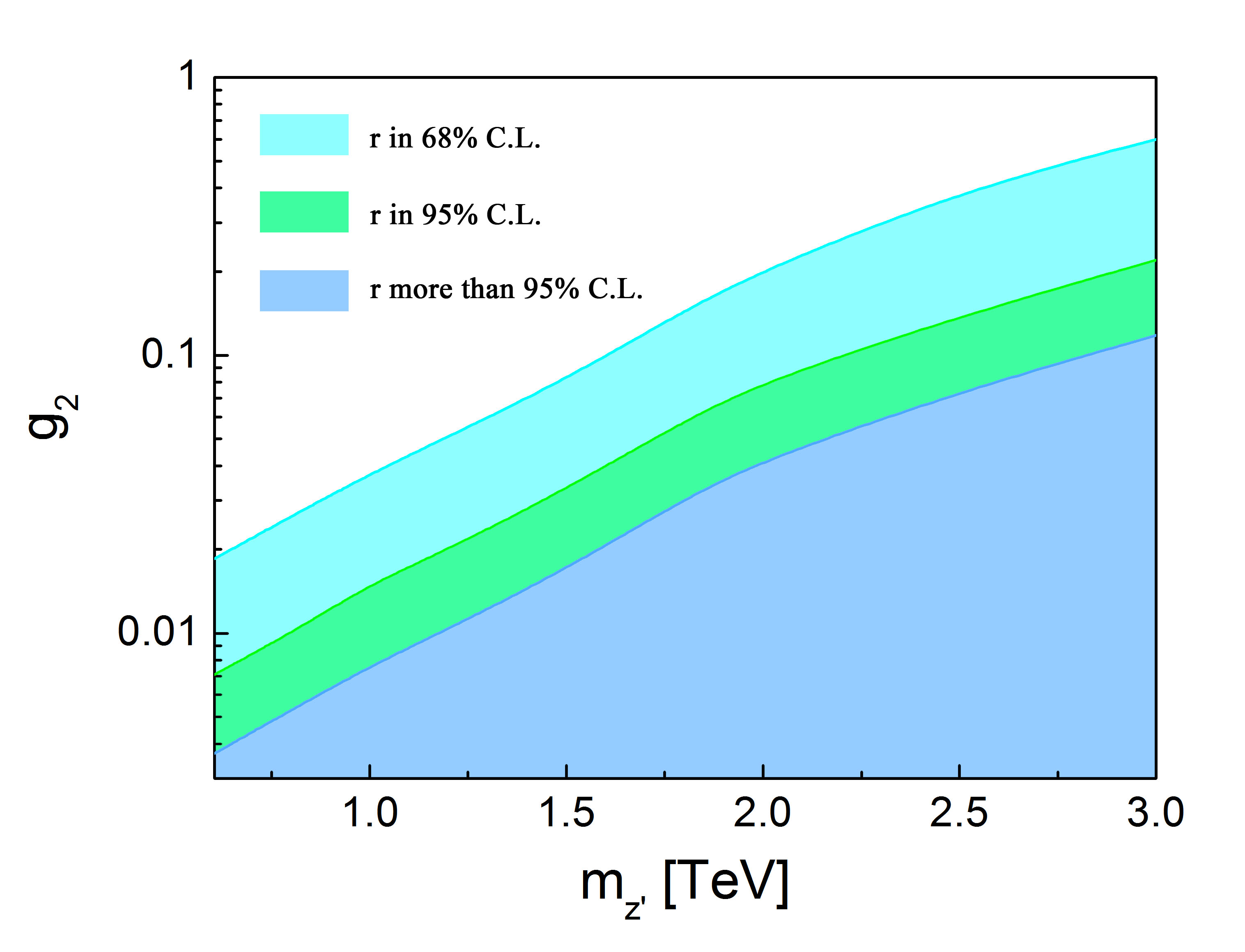}
	\caption{allowed $Z'$ mass by $7$ TeV ATLAS in $m_{Z'}$-$g_2$ plane in different charge ratio $r$ fit range corresponding $1\sigma$, $2\sigma$ and $>2\sigma$ C.L..}
	\label{Fig.mzpg2}
	\end{center}
	\end{figure}

The LHC has searched a vector resonance decaying into dilepton final states at $\sqrt{s}=7$ TeV \cite{ATLASZprime}. No statistically significant excess above the SM expectation is observed, which strictly limits $Z'$ couplings to fermions. Figure.\ref{Fig.ATLASlimit}
shows the 95\% C.L. allowed areas in $g_2$-$r$ plane with $m_{Z'}=0.8,~1.0,~1.5,~2.0$ TeV respectively. The theoretical cross sections are calculated by {\it Madgrapha5} ver1.5.12. Compared with observed limits on $\sigma(pp\rightarrow Z'\rightarrow ll)$ at ATLAS, the values of $g_2$ and $r$ can be determined with fixed $m_{Z'}$. It indicate that a light $Z'$ with enough small coupling is not eliminated.
More clearly, Fig. \ref{Fig.mzpg2} plots $7$ TeV ATLAS allowed parameter space in the plane of $Z'$ mass and coupling $g_2$, corresponding the different $r$ C.L.. A sub-TeV $Z'$ is allowed when the coupling $g_2\sim 0.01$.
	
\begin{figure}[htbp]
	\begin{center}
	\includegraphics[scale=0.10]{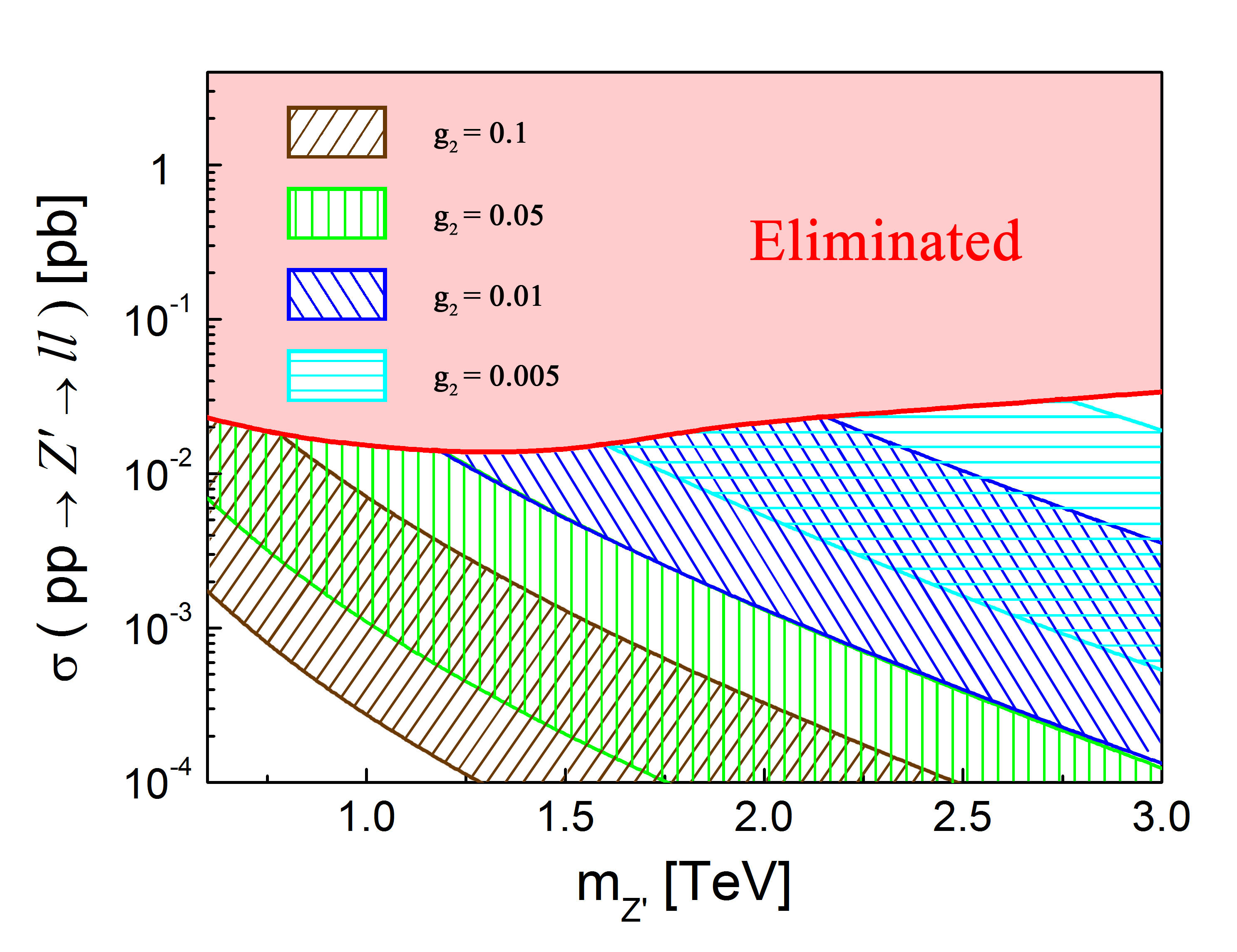}
	\caption{theoretical signal for $\sigma(pp\rightarrow Z'\rightarrow ll)$ at 14TeV LHC. The colorful massive shadow areas correspond $2\sigma$ fit range of charge ratio $r$. The red solid area on the top is eliminated by ATLAS direct detection in dilepton final states at $\sqrt{s}=7$ TeV.}
	\label{Fig.search}
	\end{center}
	\end{figure}

 A possible $\sigma(pp\rightarrow Z'\rightarrow e^+e^-,\mu^+\mu^-)$ signal in dilepton final state at $\sqrt{s}=14$ TeV LHC is also calculated by {\it Madgraph5}, which is shown in Fig.\ref{Fig.search}. For a significant coupling $g_2=0.05$, $Z'$ may still exist at about $1.2$ TeV. Even for $g_2=0.1$, $Z'$ with more that $1.6$ TeV mass is not eliminated. On the other hand, a light $Z'$ at sub-TeV would be permissible as long as the coupling $g_2$ is enough small.

\section{conclusion}

In conclusion, a model-independent effective theory for anomaly-free neutral boson has been presented. Based on electroweak precise measurements in LEP, four parameters related to $Z'$ mixings and charge assignment have been constrained. Especially, the charge ratio $r$ has range $(-2.2,2.2)$ at 95\% C.L..
To consistent with the LHC direct detection in dilepton decay channel at $7$ TeV,
the limit areas to fixed $Z'$ mass are shown in $r$-$g_2$ plane.
More clearly, we map the possible parameter space to the plane of $m_{Z'}$-$g_2$ at $2\sigma$ C.L. of $r$. in Fig. \ref{Fig.mzpg2}. The results show that a sub-TeV $Z'$ with coupling $g_2\sim 0.01$ is still not eliminated in $2\sigma$ C.L. of $r$. It suggests a prime requirement to extra vector boson in NP.  Further, a possible theoretical $\sigma(pp\rightarrow Z'\rightarrow ll)$ signal at $\sqrt{s}=14$ TeV LHC is also calculated.

\section*{Acknowledgments}
This work was supported by National Science Foundation of China (NSFC) under Grant No.11005084 and partly by the Fundamental Research Funds for the Central University. We thank Rong Li for useful discussions.


\begin{thebibliography}{1}
\bibitem{StueckelbergNote}
	When we expand covariant derivative to include Stueckelberg mechanism for $U(1)'$ mass, $\Delta_{32}$ must be non-zero to diagonalize Stueckelberg mixing. In the case, vector-type  photon coupling requires $y'^f_L=y'^f_R$ for any flavors. It yields the special charge assignment corresponding to charge ratio $y'_q/y'_u=r=1$ in anomaly-free case. So, Stueckelberg extended effective theory has the same phenomenology as $r=1$ case at the LHC.
\end{thebibliography}
\end{document}